\documentclass[twoside]{LCWS11}
\usepackage[latin1]{inputenc}
\usepackage[dvips]{graphicx,epsfig,color}
\usepackage{wrapfig,rotating}
\usepackage{amssymb,amsmath,array}
\usepackage{cite}

\begin{document}
\title{
Simulations for CLIC Drive Beam Linac} 
\author{Avni AKSOY 
\vspace{.3cm}\\
Ankara University, Institute of Accelerator Technology, Ankara, Turkey
}

\maketitle

\begin{abstract}
The Drive Beam Linac of the Compact Linear Collider (CLIC) has to accelerate an electron beam with 4.2 A up to 2.4 GeV in almost fully-loaded structures. The  pulse contains about 70000 bunches, one in every second rf bucket, and has a length of 140  $\mu$s. The beam stability along the beamline is of concern for such a high current and pulse length. We present different options for the lattice of the linac based on FODO, triplet and doublet cells and compare the transverse instability for each lattice including the effects of beam jitter, alignment and beam-based correction. Additionally longitudinal stability is discussed for different bunch compressors using FODO type of lattice.   
\end{abstract}

\section{Introduction}

CLIC~\cite{clicrep2000} is based on a two-beam scheme in which the rf power used to accelerate the main beam (at 12 GHz) is produced by a second beam (the drive beam, DB) running parallel to the main one~\cite{braun98} through so-called Power Extraction and Transfer Structure (PETS). This drive beam has a high current but relatively low energy and is decelerated for producing the rf power.

The CLIC Drive Beam Linac (DBL) will consist of about 750 structures which are low frequency (1 GHz) and will be almost fully loaded transferring more than 95\% of their input power to the beam. The average energy gain per structure will be $\Delta E \approx  3.4$ MeV~\cite{personal_com}. The initial beam energy is assumed to be $E_{0} =50$ MeV, the final beam energy $E_{f} =2.4$ GeV, the bunch charge $q =$ 8.4 nC, initial bunch length $\sigma_{z,0} = 3$ mm and the transverse normalized emittances are $\epsilon_{N,x} =\epsilon_{N,y}=50~\mu$m \cite{clicparms}. The beam pulse consists of 24 $\times$ 24  sub-trains of about 120 bunches each. The first sub-train fills odd buckets, the immediately following second  sub-train fills even buckets; this pattern is then repeated. After DBL, 24 sub-trains will be merged into a signle sub-train using delay loop (DL), combiner ring one (CR1) and combiner ring two (CR2)\cite{clicparms}. At the end of CLIC DB complex initial train with 140 $\mu$s length will be transformed to 24 trains, each has 240 ns and 100 A pulse current.

In this study, we discuss major transverse instabilities driven by wakefields in accelerating sections based on different lattice types. The linac will be seperated into two section with a bunch compressor which reduces the initial bunch length  $\sigma_{z,0} = 3$ mm to the final value of $\sigma_{z,f} = 1$ mm. In order to define longitudinal tolerances four different bunch compressor have been taken into account with neglecting the imperfections on bunch compressor sections. Also coherent synchrotron radiation has not been included. Additionally, in all calculations we have used simulation code PLACET \cite{schulte_placet} and we have taken into account only two sub-trains 15 bunches each. As it is seen later, the multi-bunch effects reach steady state condition within this length for a sub-train.

\subsection{ Layout of DBL }{\label{sec:layout}}

\begin{figure}[]
\begin{center}
 \includegraphics[width=0.8\columnwidth]{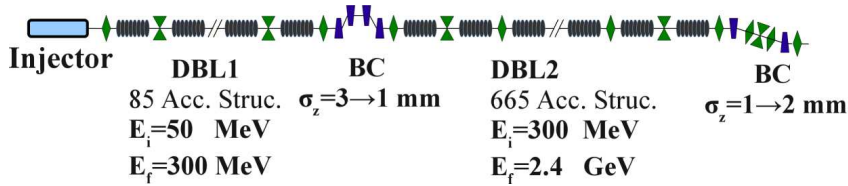}
\caption{\label{fig:dbalayout} Basic layout of CLIC drive beam linac }
\end{center}
\end{figure}

In CLIC 1\% luminosity loss requires $\delta\sigma_{\phi}\le 0.2^{\circ}$ bunch phase and $\delta\sigma_{z}\le 1~\%$  bunch length jitter in the PETS  \cite{schulte_phase,eadli}. Therefore bunch energy jitter and bunch phase-length coupling in DBL are of concern. If the full bunch compression is performed in front of PETS one needs  $R_{56} \approx -60$ cm for a chirp of 0.5\% energy spread per $3$ mm bunch length. In that case, for getting acceptable beam phase jitter one would need $3 \times 10^{-5}$ beam energy jitter. In order to avoid the strong coupling between energy jitter induced in the drive beam accelerator and beam phase jitter transformed in the bunch compressor, we propose that the bunches are accelerated to 300 MeV in first stage of drive beam linac (DBL1) and compressed from 3 mm to length of 1 mm, which is the length required in PETS, and then accelerated to their final energy of $E_{f} =2.4$GeV (see Fig. \ref{fig:dbalayout}). 

\subsection{ Accelerating Structure }

\begin{wrapfigure}{r}{0.55\columnwidth}
\centerline{\includegraphics[width=0.5\columnwidth]{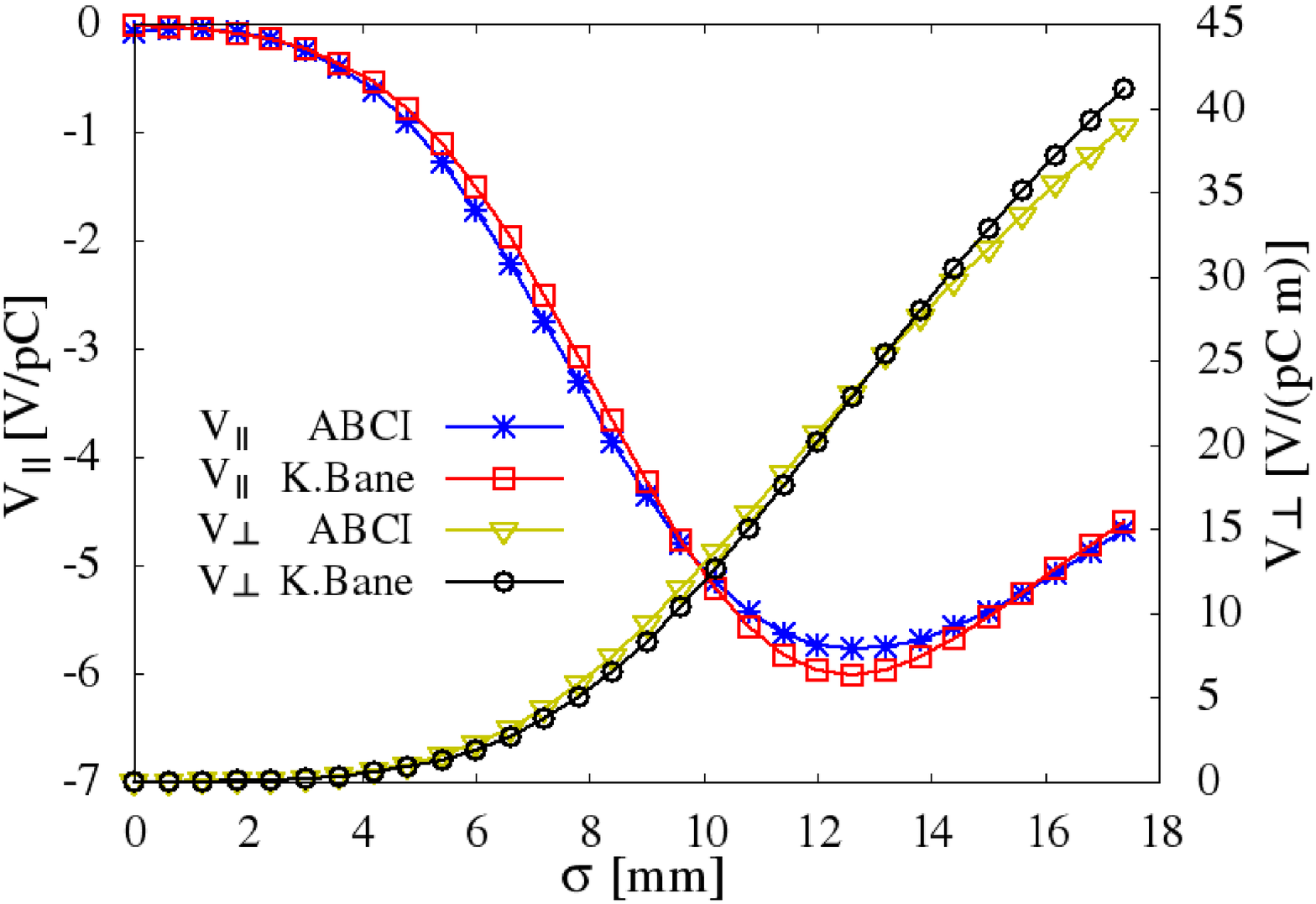}}
\caption{\label{fig:wakes} Wake potentials of a Gaussian bunch with 3 mm bunch
length }
\end{wrapfigure}
The accelerating structure, which will be fed with 15 MW input power, will consist of 19 cell in a length of 2.4 m. It will be same Slotted Iris Constant Aperture (SICA) structure like in CTF3 \cite{jensen_1q,personal_com}. Short range wake longitudinal and transverse potentials of the structure have been calculated using ABCI code \cite{abci} for a Gaussian bunch. In order to compute non-Gaussian bunch wake especially in DBL2, short range wake functions of the structure have been obtained with numerical fitting of Karl Bane's expressions \cite{bane1,bane2} to ABCI results. Figure \ref{fig:wakes} shows computed wake potentials of a Gaussian bunch using Bane's formulas and ABCI code.

The long-range transverse wakefields used in calculations have been obtained by scaling lowest four dipole modes  of 3 GHz CTF3 structure to 1 GHz \cite{ctf3des}. However the loss and damping factors used in the simulation are 50\% larger than in ref \cite{clicparms}  and almost perfect compensation of the long-range longitudinal wakefields is predicted \cite{personal_com}.

\section { Lattices }

\begin{figure}[h]
\includegraphics[width=\columnwidth]{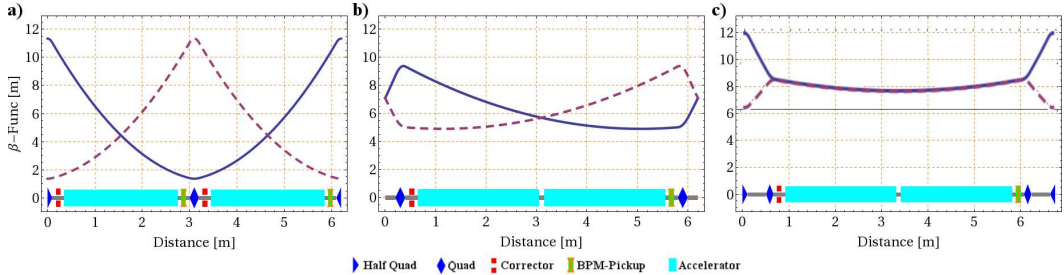}
\caption{\label{fig:lattices} Sketch of the lattice cells and betatron functions along the cells.
a)FODO, b)Doublet, c)Triplet. Strength of quadrupoles are scaled with energy. }
\end{figure}

Three different lattices were investigated with taking into account their cost. One consists of simple FODO-cells, with one structure between each pair of quadrupoles. The other lattice is based on doublets in which two structures are placed in one cell. The last one is the triplet which houses two structures similar to doublet (see Fig. \ref{fig:lattices}). Since transverse deflection caused by wakefields requires small betatron functions especially when the beam energy is small \cite{schulte_multi,avni} we have optimized lattices for minimum integration and the best phase advance along the beamline. With constant quadrupole spacing and with constant phase advance per cell, the strengths of quadrupoles reach to high values by the end of beamline (e.g. $\sim$ 0.65 T pole tip field for 22 cm quadrupole length).
In the FODO lattice, the length of each quadrupole is 20 cm, the spacing between quadrupoles is 2.9 m and the phase advance is $\mu_{x,y}= 103^{\circ}$ per cell. The doublet lattice has phase advance of $\mu_{x,y}= 58^{\circ}$, the doublet spacing is 5.4 m, the distance between two quadrupoles is 40 cm and the length of each quadrupole is 20 cm. In triplet, the distance between triplets is 5.4 m, the distances between quadrupoles in triples are 40 cm and one has $\mu_{x}= 46^{\circ}$  and $\mu_{x}= 49^{\circ}$ phase advances. For triplet, the lengths of central and outer quadrupoles are chosen 22 cm and 16 cm, respectively, also the strength of central one is larger than outer ones in order to have equal horizontal and vertical betatron functions inside the structure (see Fig. \ref{fig:lattices}). The lengths of the lattices are comparable but, obviously, triplet would have more cost due to one more quadrupole for each accelerating structure.

In all following calculations tracking has been started and finished at the middle of the distance between two quadrupoles for FODO, it is middle of the distance between doublets and triplets for other relevant types of lattices. This choice gives availability to align all quadrupoles in misalignment studies since the alignment is performed respect to the following BPM after quadrupole. However the minimum transverse acceptances are 5.33$\sigma$, 6.20$\sigma$ and  5.04$\sigma$ for the FODO, doublet and triplet lattice, respectively.

\section { Transverse Beam Jitter}{\label{jitter}}

Since we can not estimate the transverse jitter of the incoming beam, only the jitter amplification is calculated. The normalized amplification factor $Amp$ for a slice, that has $\Delta x_{0}$ initial offset, is defined as: 
\begin{equation}{\label{equ:2} }
Amp_x = \frac{1} {x_{N}(0)}
\sqrt{ x_{N}^2(L) + {x'}_{N}^2(L)  }
\end{equation}
Here, $L$ is length of beamline, $x_{N}(0)$, $ x_{N}(L)$ and  ${x'}_{N}(L)$  are initial position, final position and final angle of the center of the slice in normalized coordinates, respectively. Equivalently one can define $Amp_y$ and the maximum amplification factor $Amp_{max}$ is the maximum of $Amp_x$ and $Amp_y$ over all slices. For a slice with nominal energy and without wakefield effects, one has $Amp_{x,y}=1$. In order to check the amplification of bunches in a train, we use the method in Ref. \cite{schulte_multi,avni}. 

\begin{figure}[h!]
\includegraphics[width=\columnwidth]{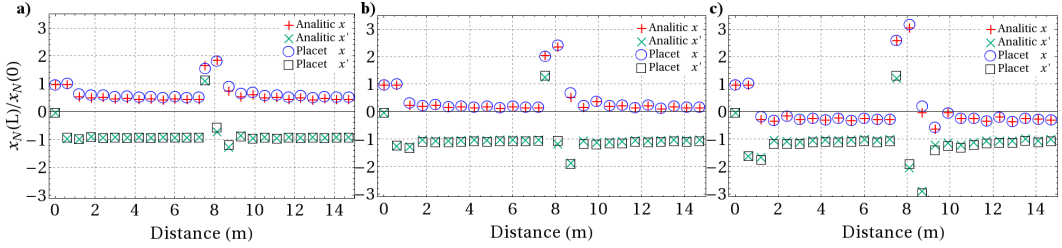}
\caption{\label{fig:pointamp} The normalized amplitudes of the point-like bunches with constant charge at the end of perfectly aligned DBL for an offset incoming train, a)FODO, b)Doublet, c)Triplet.  }
\end{figure}

The impact of an initial offset of a train of point like bunches with constant charge is shown in Fig. \ref{fig:pointamp}. We assumed all bunches have the nominal Twiss parameters and the same initial offset at the entrance of beamline. In calculation we have used two sub-trains of 15 bunches each and for checking the worst case we take into account full bunch charge at sub-train switching point.  As it can be seen on the figure the amplification of bunches of a single sub-train reaches steady state rapidly within this sub-train length and the agreement between the simulation and the simple analytic model is very good. Since the distance between bunches at switching point from odd buckets to even (or v.v.) is half of the others, the amplification at that point is slightly high due to strong kick caused by closer bunches. FODO lattice compensates transverse deflections and worst one occurs on triplet. The maximum amplification factor, $Amp_{max}$, for point-like bunch case for FODO, doublet and triplet lattices are 2.03, 2.65 and 3.67, respectively. In case of half bunch charge at sub-train switching point the amplification factors will decrease to 1.41, 1.77 and 2.34 for FODO, doublet and triplet lattices, respectively \cite{avni}.

\begin{figure}[h!]
\centerline{\includegraphics[width=0.7\columnwidth]{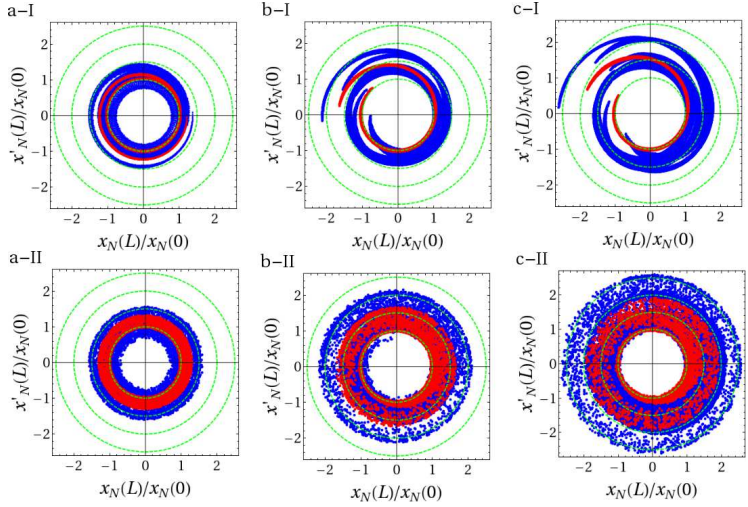}}
\caption{\label{fig:multiamp} The normalized final offset of the bunches with constant charge at the end of the DBL1 (I, top) and DBL2 (II, bottom) for an offset incoming train, a)FODO, b)Doublet, c)Triplet. Group of red dots shows first bunch and blue dots show trailing bunches. }
\end{figure}

For multi-particle case of bunches, there is additional transverse kick due to short range wake fields. On the other hand the energy difference of particles within a bunch will force them to advance  in  phase  with respect to the reference one, thus some compensating of the kicks of long range wakefields occurs \cite{mosnier,bane1986}. Therefore the amplification factor will not be as high as point-like bunches case. Fig. \ref{fig:multiamp} shows PLACET results for final offset of a train at the end of perfectly aligned DBL1 and DBL2. Similar to point-like bunches case, all bunches have nominal Twiss parameters at the entrance of beamline and train consists of two sub-trains 15 bunches each. Switching point from even to odd buckets the bunches are kicked significantly; the maximum amplification for FODO, doublet and triplet lattices are 1.55, 2.15 and 2.70, respectively. Without knowledge of the acceptance downstream and the size of the incoming beam jitter, it is not possible to decide whether the amplification is acceptable. For all types of lattices within the linac that has $5\sigma$ minimum acceptance, even a large incoming jitter of $\Delta x_0=\sigma$ does not lead to beam loss.

\section { Alignment }

All elements on beamline may be scattered around a straight line. To align the beamline and compute the emittance growth caused by the miss-alignment, two different routines based on the beam have been taken into account. First one-to-one correction: each quadrupole is moved transversely in order to bring the average beam position to zero in the beam position monitor (BPM) located after quadrupole. Second wakefield-free steering:  two or more beams with different energy and charge from the nominal one are steered to follow the same trajectory in order to remove dispersion and wakefield effect from the lattice during one-to-one correction is applied to nominal beam \cite{wkfresteering}.

In order to have better comparison between lattices we have used one BPM after each quadrupole at an appropriate location and have simulated 100 different beamlines, the elements of which are scattered with a normal distribution. In calculation following assumptions have been considered:
\vspace{-\topsep}
\begin{itemize}
 \item all quadrupoles have $\sigma_{x,y}=300$ $\mu$m position errors $\sigma_{x',y'}=300$ $\mu$rad angle errors and $\sigma_{\theta}=1$ mrad roll errors;
\vspace{-\topsep}
\item all BPMs and accelerating structures have only $\sigma_{x,y}=300$ $\mu$m position errors;
\vspace{-\topsep}
\item the beamline on bunch compression section is perfectly aligned;
\vspace{-\topsep}
\item accelerating structures are perfectly straight (no tilting effect);
\vspace{-\topsep}
\item the beam is injected without any offset to DBL1 and DBL2;
\vspace{-\topsep}
\item the resolution of BPMs are $10$ $\mu$m;
\vspace{-\topsep}
\item each of two test beams used for wakefield-free steering consists of single bunch and  they have $E_{in,1}$ = 40 MeV and $E_{in,2}$ = 60 MeV initial energies, $Q_1$ = 9 nC and $Q_2$ = 8 nC charges, $V_{1}$ = 0.93$V_{0}$ and $V_{2}$ = 1.05$V_{0}$ accelerating gradients, respectively, where $V_{0}$ is nominal gradient for actual beam.
\end{itemize}

\begin{wrapfigure}{r}{0.65\columnwidth}
\includegraphics[width=0.6\columnwidth]{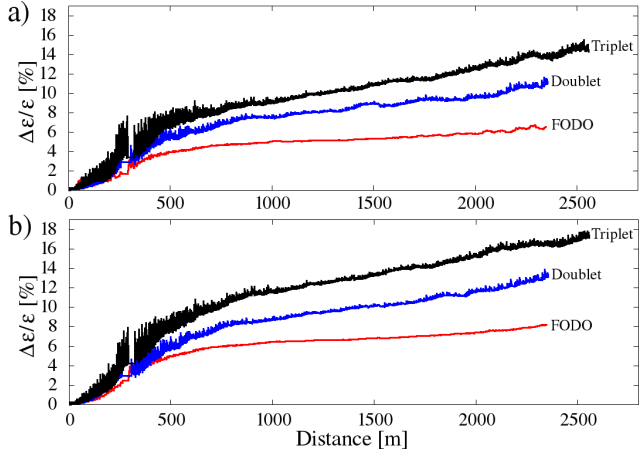}
\caption{\label{fig:emitt} Emittance growth along the beamline. a)wakefield-free steering, b)
one-to-one correction}
\end{wrapfigure}

Fig. \ref{fig:emitt} shows PLACET results for emittance growth along the beamline based on lattices considered. The growth is quite small for all lattice types and for both correction routines. Since the FODO type of lattice has weakest quadrupoles, the growth on it is smallest and is about 5\%. The emittance growth would be higher due to the fact that the bunch compression section will also be misaligned and the beam will have offset at the entrance of DBL2. However the static imperfections errors given above could be improved somewhat if necessary. 

\section {Impact of the Energy and Gradient Errors}

Although CLIC DB has very tight tolerances concerning error of incoming beam energy and structure gradients \cite{schulte_phase}, during commissioning large energy and gradient errors may occur. Any error of incoming beam energy or variation of the gradient will lead quadrupole strengths not to be adapted to the beam energy. These situations can cause beam amplification to grow, eventually, beam losses especially in DBL1 where the beam energy is low. On the other hand one can also define the tolerances for the RF and incoming beam for the real operation according the acceptable errors \cite{schulte_phase} at the end of DBL using different bunch compressors.

\subsection{Transverse Stability}

\begin{figure}[h!]
\includegraphics[width=\columnwidth]{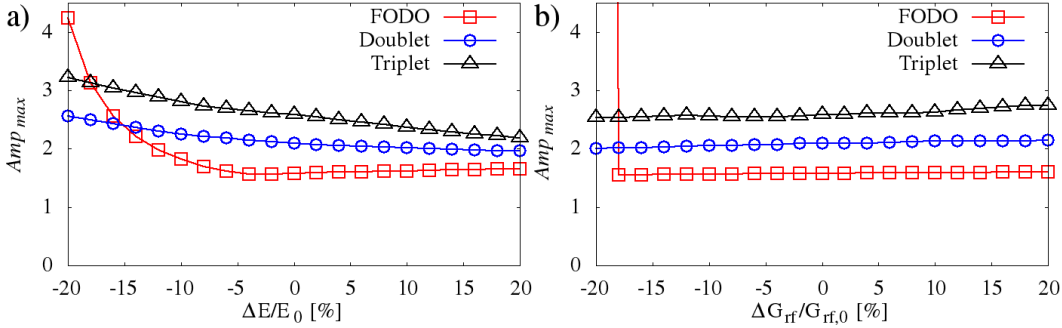}
\caption{\label{fig:eg_amp} Amplification of the beam in DBL1, a) for different initial beam energy, b) for different accelerating gradient}
\end{figure}

In order to check the amplification we have simulated train of two sub-trains 15 bunches each on perfectly aligned beamline of the lattice types considered. We assume the beam has the nominal Twiss parameters and offset of $\Delta x_0=\Delta y_0=\sigma$ at the entrance of beamline. Fig. \ref{fig:eg_amp} a shows the amplification as a function of the deviation from the nominal initial beam energy (a) and the nominal accelerating gradient (b). As it was discussed in section \ref{jitter} minimum amplification is obtained with FODO lattice for nominal case. FODO lattice is more sensitive to initial beam energy variations than the others especially while the beam energy is much lower than the nominal value. The gradient variation more or less does not change the amplification for all lattices and doublet lattice seems more stable for both energy and gradient variations.  Gradient errors below -15\% cause amplification to increase rapidly for FODO type of lattice. This result can be explained as: towards the end of DBL1 the beam, that is accelerated to lower energy than the nominal by low gradient structures, will be over focused by quadrupoles which are adapted to nominal energy. Thus, the betatron functions will grow rapidly  as a result of large phase advance. This situation is not the same for doublet and triplet lattice because of small phase advance per cell. 

Another crucial subject would be the acceptance of the lattice in DBL1. The acceptance in normalized coordinates defined as, the beam any particle of which has initial positions $x_0$, $x'_0$, $y_0$ and $y'_0$, that fulfills

\begin{equation}{\label{equ:5} }
Ar_N \geq  \sqrt{ {x}_N^2 + {x'}_{N}^2 + y_N^2 + {y'}_{N}^2 }
\end{equation}
will pass through the accelerator. In simulations, we have considered two cases: perfectly aligned beamline and beamline the elements of which are scattered with a normal distribution of $\sigma = 300~\mu$m. For miss-aligned case we have simulated 20 different machines and applied one-to-one correction. Additionally, for both case we assumed the beam which has nominal Twiss parameters, has offset $\Delta x_0=\Delta y_0=500 \mu$m at the entrance of beamline and computed the unnormalized acceptance. However we have neglected initial beam angle with expecting that it will transform same as offset under first order approximation.

\begin{figure}[h!]
\includegraphics[width=\columnwidth]{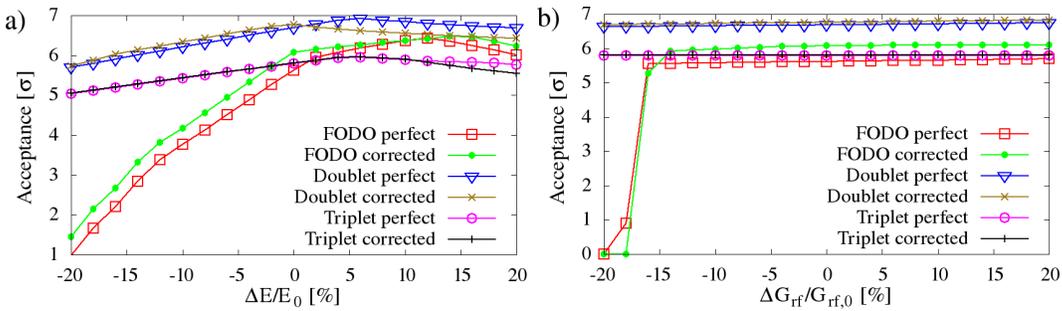}
\caption{\label{fig:egaccpt} Acceptance of the lattices, a) for different initial beam energy, b) for different accelerating gradient}
\end{figure}


Fig. \ref{fig:egaccpt}  shows the acceptance of the lattices as a function of initial beam energy error (a) and accelerating gradient error (b). The acceptance is highest for doublet and it is  less sensitive to the energy errors in doublet and triplet. For FODO the acceptance is reasonable around nominal beam energy, but it is very sensitive to the energy errors due to the same reason explained in amplification calculation. For all lattices the size of the acceptance for perfect and miss-aligned machines are close to each other. All lattices are less sensitive to the gradient error and the acceptance is highest for doublet lattice  similar to energy error case.  For FODO lattice beam loss starts when the accelerating gradient is below -15\% of the nominal gradient.

\subsection{Longitudinal Stability }

As it has been discussed in section \ref{sec:layout}, compressing the bunch before the main part of the acceleration  one can afford having a strong bunch energy chirp  and small $R_{56}$, thus a weak coupling between beam energy jitter and beam phase jitter can be obtained.  Assuming additional improvement by factor 10 for the tolerance of the phase using feed forward system before PETS the energy error in DBL1 relaxes up to $1 \times 10^{-3}$ \cite{schulte_phase}. In the second stage of the drive beam linac (DBL2), the large relative energy spread will be reduced below 0.4\% which is acceptable in the PETS. In order to reduce significant impact of coherent synchrotron radiation, the bunches are uncompressed to 2 mm before they enter the delay loop and re-compressed behind the combiner rings to the final required length of 1 mm. To avoid an energy jitter from DBL2 turning into beam phase jitter the sum of all $R_{56}$ of all elements after DBL has to be zero. 

Under first order approximation the error of beam phase ($\delta \sigma_{\phi}$)  or error of bunch length ($\delta \sigma $)   of a beam in a magnetic chicane will be proportional to jitter of the beam energy ($\delta \sigma  \approx R_{56} \delta E $).  Several quantities such as error of charge ($\delta Q$), incoming beam phase ($\delta \sigma_{in}$)) and energy ($\delta E_{in}$)  of incoming bunch or error of phase ($\delta \phi_{RF}$) and gradient ($\delta G$) of linac can cause jitter on relative energy spread.  The bunch compressor should compensate large errors of these quantities as well as errors caused by beam loading. Currently we have taken into account single bunch case and studied four types of compressors which have -10 cm, -12 cm, -14 cm and +26 cm compressing factors in order to define longitudinal tolerances with neglecting the imperfections on bunch compressor sections.

\begin{figure}[h!]
\includegraphics[width=\columnwidth]{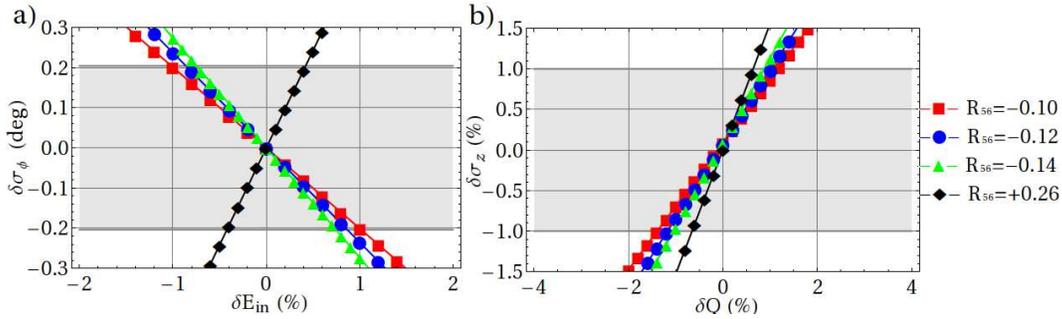}
\caption{\label{fig:bef_sz} Energy error of incoming bunch vs. phase jitter of outgoing bunch (a), and charge error of incoming bunch vs. bunch length variation of outgoing bunch (b). Gray areas show acceptable tolerances. }
\end{figure}

Left hand side of  Fig. \ref{fig:bef_sz} shows phase jitter of the outgoing bunch as a function of deviation from nominal initial beam energy.   As it can be seen on the figure largest incoming beam energy error is accepted by the bunch compressor that has $R_{56}$=-10. Right hand side of the Fig. \ref{fig:bef_sz} shows bunch length variation as function of the deviation from the nominal bunch charge. Similarly largest charge error (current error) is  compensated by by the bunch compressor that has $R_{56}$=-10 while smallest errors are accepted by the compressor with $R_{56}$=+26 cm.

\begin{figure}[h!]
\includegraphics[width=\columnwidth]{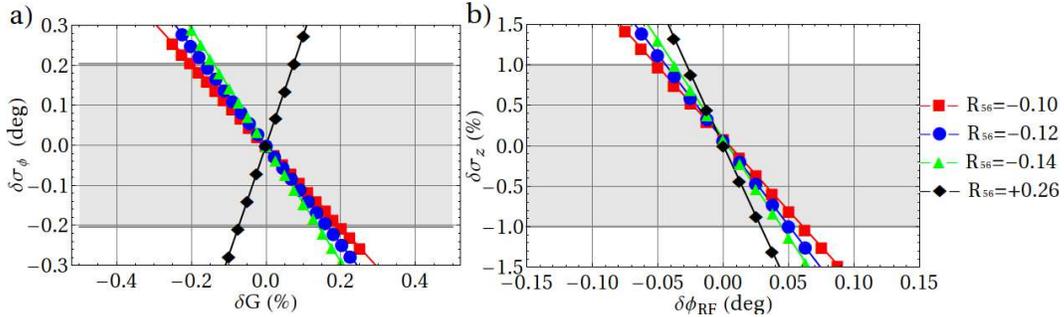}
\caption{\label{fig:lf_sz} Gradient error of linac vs. phase jitter of outgoing bunch (a), RF error of linac vs. bunch length variation of outgoing bunch (b). Gray areas show acceptable tolerances.}
\end{figure}

 Left hand side of the Fig. \ref{fig:lf_sz} shows phase jitter of the outgoing bunch as a function of linac gradient error in DBL1 and right hand side shows bunch length variation as function of phase error of linac in DBL1. Similarly to initial bunch energy and charge error cases largest errors are accepted by the bunch compressor that has $R_{56}$=-10cm.

\section { Conclusion }

The lattice has to prevent a large amplification of any transverse jitter of the incoming beam.  It should also have a large energy acceptance and allow easy correction of static errors of the beam line. Three type of lattice have been studied for finding a compromise between lattices for the CLIC DBL. 

The calculations show that if one uses FODO type of lattice the effects of transverse wakefields will be significantly smaller than using doublet or triplet types. For both alignment routines, triplet type of lattice gives largest emittance growth. Although FODO and doublet type of lattices have same number of quadrupoles,  FODO  gives smallest emittance growth. On the other hand, smallest sensitivity to energy errors can be obtained with doublet type of lattice while FODO type of lattice yields largest. Nevertheless, too large errors should not be important since the energy of the beam has to be controlled very accurately because of the tight tolerance of energy error in bunch compressor.   

The FODO lattice has as many magnets as the doublet solution but fewer than the triplet design. It performs best in terms of jitter amplification and emittance growth in presence of static imperfections. The energy bandwidth is smaller than it is in the other designs but we consider the first two points more important. In particular the jitter amplification is very important since this can lead to losses further downstream. This will be a change compared to the CTF3 design, which is based on triplets \cite{ctf3des}. 

FODO lattice will bring out other advantages such as easy operation. One can also consider using four structures in one FODO-cell after 1.5 GeV. For example, if one uses 30\% weaker and 40\% shorter quadrupoles after 1.5 GeV the integral increases from 11.28 m$^2$/MeV to 12.439 m$^2$/MeV and beamline shortens from 2.35km to 2.21km. In that case the maximum amplification factor of a train of point-like bunches will be 2.22 which is still better than doublet case mentioned above \cite{avni}.

One can also improve the triplet lattice with using single accelerating structure in one triplet-cell as it is in CTF3 design. In that case the maximum amplification factor of a train of point-like bunches can be calculated as 2.58 for the constant bunch charge condition considered in section \ref{jitter}. On the other hand using single structure per cell would increase the length of beamline from 2,55km to 3.25km.

\begin{wraptable}{l}{0.5\columnwidth}
\centerline{
\begin{tabular}{|l|c|c|}
\hline
{\bf Parameter}  & {\bf Value} &  {\bf Unit} \\ 
\hline
 Initial energy error & 1 & \% \\ 
\hline
 RF power error & 0.2 & \% \\ 
\hline
 Beam current error & 0.1 & \% \\ 
\hline
 RF phase error & 0.05 & deg \\ 
\hline
\end{tabular}
}
\caption{The tolerances in DBL1.}
\label{tab:limits}
\end{wraptable}

As it can be seen on figures \ref{fig:bef_sz} and \ref{fig:lf_sz} above, larger longitudinal errors can be compensated by chicane that has $R_{56}= -10$ cm. In order to compress the bunches from 3 mm to 1 mm with that bunch compressor one can easily calculate a chirp about 2\% energy spread per $3$ mm bunch length which means the RF will be operated at about 25$\circ$ off-crest phase.  Thus relative gradient reduction comes out for this compressor while the compressor with $R_{56}= +26$ cm allows RF to be used more efficient. Calculated longitudinal tolerances of DBA1 is summarized as in Table  \ref{tab:limits}. Additionally to the table the tolerance of accelerating power amplitude will be half of klystron power and the phase error of the incoming beam  will be two times of the RF phase error.  

\section{Acknowledgments}

Author would like to thank Dr. Daniel Schulte for his support during the study of beam dynamics for CLIC Drive Beam Linac.

This work is partially supported by State Planning Organization (SPO) of Turkey  and CERN.

\bibliography{a_aksoy_LCWS11_clic_DB_linac_sim_v1}

\end{document}